\documentclass{article}

\usepackage[preprint]{neurips_2022}

\usepackage[utf8]{inputenc} 
\usepackage[T1]{fontenc}    
\usepackage{hyperref}       
\usepackage{url}            
\usepackage{booktabs}       
\usepackage{amsfonts}       
\usepackage{nicefrac}       
\usepackage{microtype}      
\usepackage[table,xcdraw]{xcolor}
\usepackage{graphicx}       
\usepackage{threeparttable} 
\usepackage{multirow}
\usepackage[font=small,labelfont=bf]{caption}
\usepackage{float}

\title{WuYun: Exploring hierarchical skeleton-guided melody generation using knowledge-enhanced deep learning}
\author{%
	{Kejun Zhang$^{1,2,3,*}$, \quad Xinda Wu$^{1,}$\thanks{Equal contribution.} \ , \quad Tieyao Zhang$^1$, \quad Zhijie Huang$^1$, \quad Xu Tan$^4$,}\\
	{\textbf{Qihao Liang$^1$, \quad Songruoyao Wu$^1$, \quad Lingyun Sun$^{1,2,}$\thanks{Corresponding author.}}}\\
	$^1$College of Computer Science and Technology, Zhejiang University, China.\\ 
        $^2$Alibaba-Zhejiang University Joint Institute of Frontier Technologies, China.\\ 
        $^3$Innovation Center of Yangtze River Delta, China.\\ 
	$^4$Microsoft Research Asia\\
        \texttt{\{zhangkejun, wuxinda, kreutzer0421, zj\_huang, } \\
        \texttt{qhliang, 12221193, sunly\}@zju.edu.cn} \\
	\texttt{xuta@microsoft.com} \\ 
}

\begin{document}

\maketitle

\begin{abstract}
  Although deep learning has revolutionized music generation, existing methods for structured melody generation follow an end-to-end left-to-right note-by-note generative paradigm and treat each note equally. Here, we present WuYun, a knowledge-enhanced deep learning architecture for improving the structure of generated melodies, which first generates the most structurally important notes to construct a melodic skeleton and subsequently infills it with dynamically decorative notes into a full-fledged melody. Specifically, we use music domain knowledge to extract melodic skeletons and employ sequence learning to reconstruct them, which serve as additional knowledge to provide auxiliary guidance for the melody generation process. We demonstrate that WuYun can generate melodies with better long-term structure and musicality and outperforms other state-of-the-art methods by 0.51 on average on all subjective evaluation metrics. Our study provides a multidisciplinary lens to design melodic hierarchical structures and bridge the gap between data-driven and knowledge-based approaches for numerous music generation tasks.
\end{abstract}

\section{Introduction}
Automatic music generation is one of the popular multidisciplinary research topics in generative art and computational creativity (1), which has achieved revolutionary advances in various artificial intelligence-generated content applications by utilizing deep learning techniques (2, 3), including interactive music production collaboration tools (4, 5), video background music generation (6), music education (7), and music therapy (8). As one of the crucial components of music generation, melody generation has drawn much attention from both the academic and industrial fields. Although melodies appear to be a simple linear succession of notes unfolding over time, the organizational structure of the melodic notes is hierarchical, like a tree resulting in intricate long-distance dependencies (9, 10). Hence, the complex long-distance dependencies make it difficult for neural networks to discover and learn the hierarchical structure relationships among these musical elements and generate long-term structured melodies. In recent years, language models in natural language processing (NLP) have been employed to capture long-distance dependencies for structured melody generation with the advantages of an easy-to-use end-to-end deep learning framework, effective representation learning, and arbitrary sequence length generation. Their powerful ability to automatically learn the latent knowledge from big data, without explicitly codifying the domain-specific rules, has been proved and applied in multiple disciplines (11–14).

Numerous specialized architectures of the language model for music generation have demonstrated promising performance in generating long-range coherent melodies, including effective attention mechanisms (15, 16), enhanced memory networks (17–19), large-scale deep neural networks (20), and explicit musicality regularization (21). Furthermore, various MIDI-derived symbolic music representation methods designed auxiliary musical spatiotemporal symbols (e.g., BAR, POSITION, and CHORD) for the input symbolic music data to help music generation models learn the long-distance dependencies better, longer, and faster (17–19, 22, 23). However, the scarcity of publicly available melody data limits the usage of the power of language-based music generation models. Moreover, the process of melody generation still lacks controllability. These models are trained in the dominant end-to-end and data-driven learning paradigms, which optimize the network's large-scale parameters via learning to map the input data to output data, thus occasionally resulting in excessive repetition or boring sounds in the generated music (21).

Recent studies used a deep learning-based hierarchical generation strategy to first hallucinate or predict the object’s structure and then use it to constrain downstream generation tasks (e.g., protein, font, or music) (24–33), which enables the neural networks to learn from the limited data far more efficiently and improves the controllability of the generation process. For structured melody generation, some scholars first generate a melody's hierarchical music structure representation (31) or bar-level musical structure relationship graph (32, 33) and then generate melodies conditioned on the generated parallel structure information as additional knowledge. Such a strategy requires recognizing the group structure of the musical syntax in a melodic surface (e.g., phrases and sections) to extract music features for building a structure generation model. Nonetheless, inadequate music structure boundary detection algorithms hinder the extraction of accurate melodic group structure. Conversely, little attention has been paid to the organizational logic of the deep structure beneath the melodic surface, organized by different levels of structural importance among various musical events (34–36) with the potential to enhance structured melody generation. Typically, the majority of existing melody generation methods for pursuing long-term structure follow an end-to-end left-to-right note-by-note generative paradigm and treat each note equally. So far, however, there is still an insufficient investigation into an alternative order of melody generation and the difference in the relative structural importance among musical notes.  

In this study, we propose WuYun, a hierarchical skeleton-guided melody generation architecture based on knowledge-enhanced deep learning that incorporates the melodic skeleton as deep structural support to provide explicit guidance on the development direction of melody generation (Fig. 1A). WuYun follows the hierarchical organization principle of structure and prolongation (35, 37), thus dividing traditional single-stage end-to-end melody generation into two stages: melodic skeleton construction and melody inpainting (Fig. 1B). At the stage of melodic skeleton construction, we first extract the most structurally important notes in a musical piece from rhythm and pitch dimensions as melodic skeletons on the basis of the music domain knowledge. We then train an autoregressive decoder-only Transformer-based network (38) on the collected melodic skeleton data to construct new melodic skeletons (Fig. 1C, a). We treat the melodic skeleton as the underlying framework of the final generated melody. At the stage of melody inpainting, we adopt a Transformer encoder–decoder architecture (39) to elaborate the melodic skeleton into a full-fledged melody by encoding the melodic skeleton as additional knowledge into the decoder to guide the melody generation process (Fig. 1C, b). To prove the effectiveness of the architecture, we evaluate WuYun on a publicly available melody dataset. Experimental results show that the generated melodic skeleton has comparable quality with the real one extracted by our proposed melodic skeleton extraction framework. The hierarchical skeleton-guided melody generation architecture effectively improves generated melodies’ long-term structure and musicality and outperforms other state-of-the-art methods by 0.51 on average on all subjective evaluation metrics.

\begin{figure}
\centering
\includegraphics[width=1\textwidth]{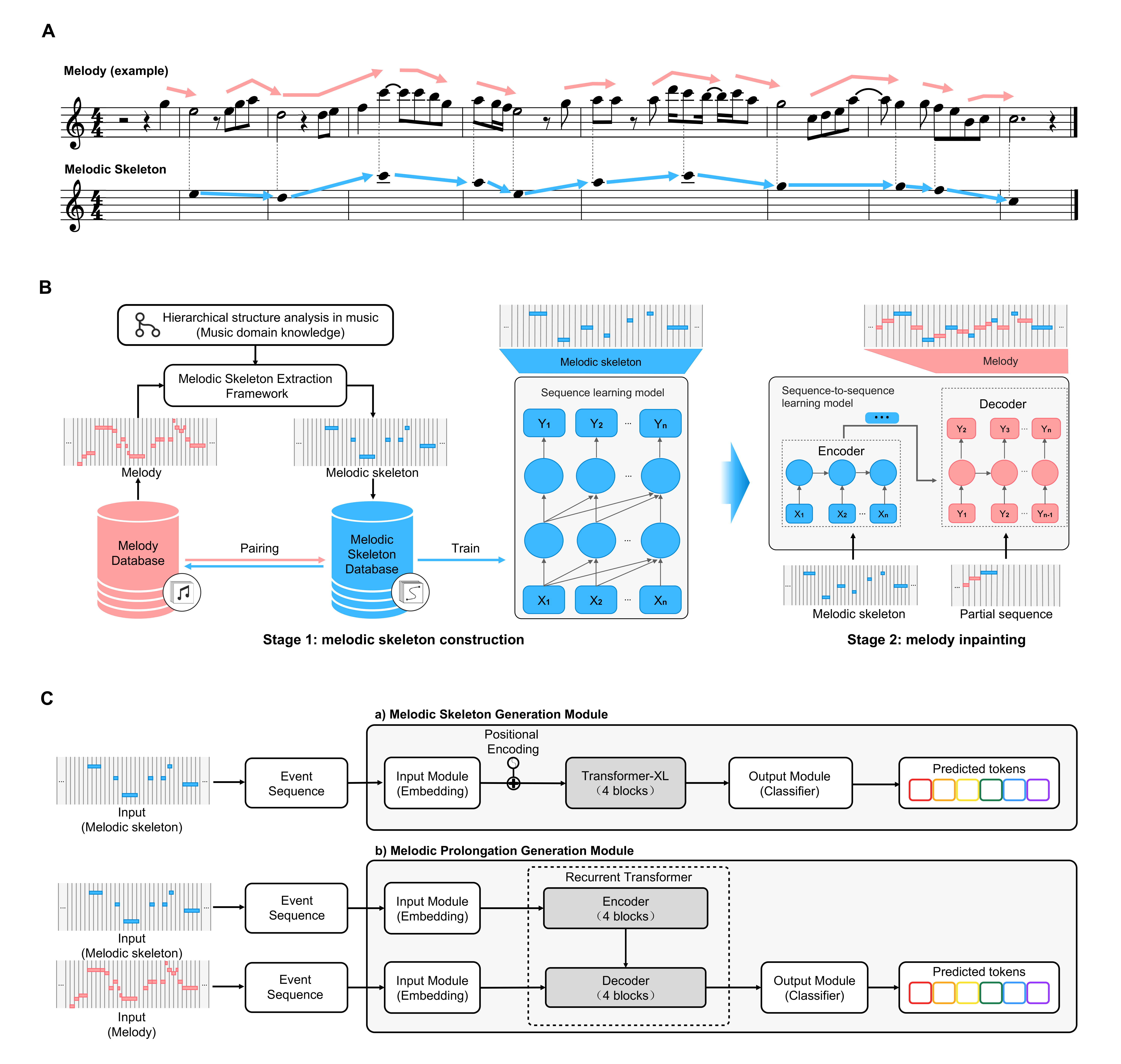}
\caption{\textbf{Architecture of WuYun.} (\textbf{A}) The first eight bars of the melody of “Hey Jude” from The Beatles (excluding anacrusis). The upper part of the figure shows the basic shape of the melodic motion, and the low part of the figure shows the melodic skeleton in the rhythm dimension. Every melody has an underlying melodic skeleton that provides structural support and connections among musical elements to guide the melodic motion. (\textbf{B}) Hierarchical melody generation process. WuYun divides the melody generation process into melodic skeleton construction and melody inpainting stages following the hierarchical organization principle of structure and prolongation. At the melodic skeleton construction stage, the melodic skeleton extraction framework is proposed to extract the melodic skeleton in the rhythm and pitch dimensions by the hierarchical structure theory from music domain knowledge. A neural network for sequence learning trained on melodic skeletons can generate novel ones. At the melody inpainting stage, another neural network for sequence-to-sequence learning would fill the generated melodic skeleton into a full-fledged melody. (\textbf{C}) Architecture details of WuYun. WuYun is composed of a melodic skeleton generation module and a melodic prolongation generation module; the former is used for the melodic skeleton construction stage, and the latter is used for the melody inpainting stage with the guidance of the melodic skeleton.}
\label{fig_architecture}
\end{figure}

\section{Result}

\subsection{Hierarchical organization principle of structure and prolongation}
Most AI artistic generative models differ significantly from humans in their artistic creation process, especially in music generation. For example, a typical music generation model generates music content sequentially from left to right at once (40). However, human artworks tend to develop iteratively from a basic underlying idea or structure through elaboration, expansion, and individual shaping. Human artistic creation follows an age-old fundamental principle of creative thinking, namely, structure and prolongation, which has significantly contributed to human thinking and creativity. In the art of music, this principle governs the underlying logic in musical composition and makes musical reasoning and explanation comprehensible and acceptable (37). It conforms to the brain’s cognitive processing mechanism of structurally organizing sequential information (41–44), which makes the brain encode and process information more efficiently and improves musical memories (45, 46). For example, musicians use this principle, consciously or unconsciously, to study, organize, and perform their musical works.

The hierarchical structure is a key feature of the tonal musical syntax system, where musical elements are almost always hierarchically organized by strict rules at a fundamental level rather than unlimited creative expression (36). Some researchers have investigated the patterns of structural organization and generalized them into music theories regarding the hierarchical structure in music from the perspective of the structure and prolongation principle. Schenker (47) was the ﬁrst to introduce this principle to describe the musical structure in a hierarchically organized way. The central idea of Schenkerian theory about the hierarchical structure is that some musical events are elaborated by other musical events in a recursive and embedded fashion (9, 34). That is, not all musical events are equally important. Some musical events have structural importance as stable factors in music, whereas others are more decorative as dynamic factors. Therefore, Schenker proposed different levels of structure hierarchy to organize tonal music and analyze its motion. Based on Schenker’s ideas, the generative theory of tonal music (GTTM), proposed by Lerdahl and Jackendoff (34), is one of the most inﬂuential theories in current music theory and music psychology. GTTM provides a systematic analysis and description of the hierarchical structure in music. Based on the listeners’ perception of tonal music, GTTM lists four hierarchical structure relationships from rhythm and pitch dimensions: grouping structure, metrical structure, time span reduction, and prolongational reduction. The term “reduction” refers to the stepwise reduction of less important musical events from the musical surface, revealing the underlying framework or skeleton that plays an essential role in the music’s qualities and developmental direction. In summary, under the surface of the music, musical events are hierarchically organized based on the structural stability in rhythm and pitch dimensions (48, 49).

Inspired by the iterative mode of human composition guided by the principles of structure and prolongation, the whole process of melody creation can be seen as progressively filling individual decorative notes among the melodic skeleton; it is an effective modern composition technique that perfectly combines rules and composers’ personality (35). This composition technique has been developed and applied in music teaching for a long history. In the following, we elaborate on the melodic skeleton extraction framework from rhythm and pitch dimensions and introduce the design of WuYun melody generation architecture that first constructs the melodic skeleton and then completes the melody instead of sequentially generating a melody note-by-note at once. The manner of WuYun’s melody generation process is more musically meaningful than the dominant end-to-end left-to-right note-by-note melody generation paradigm.

\begin{figure}
\centering
\includegraphics[width=1\textwidth]{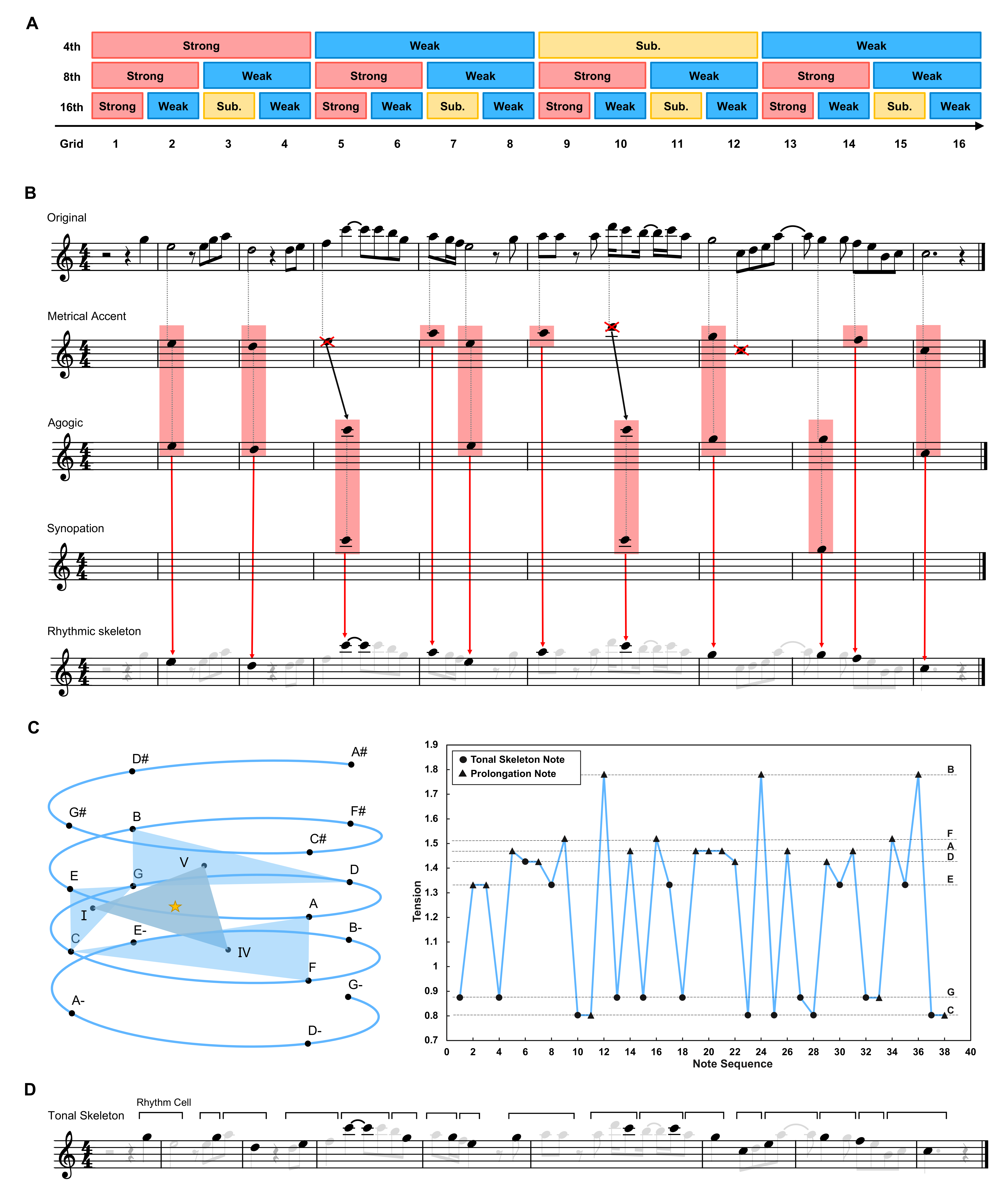}
\caption{\textbf{Melodic skeleton extraction framework.} (\textbf{A}) Rhythm pattern of strong and weak beat distribution in the 4/4 time signature with different note resolutions. (\textbf{B}) Rhythmic skeleton extraction. The rhythmic skeleton consists of the metrical accents, agogic accents on metrical accents, and agogic accents on syncopations in each measure. (\textbf{C}) Illustration of the tension measure in the pitch class helix of the spiral array with a C major chord. The right part of the figure presents the tension value of the notes in the first eight bars of the melody of “Hey Jude.” (\textbf{D}) Tonal skeleton extraction. The tonal skeleton consists of the notes with the minimum tension value in the rhythm cell.}
\label{fig_melodic_skeleton}
\end{figure}

\subsection{Melodic skeleton extraction framework}
Music theories present that there is an underlying identifiable framework beneath the melody surface called the melodic skeleton (35, 50). The melodic skeleton is composed of certain notes, which sound more structurally important from rhythm and pitch dimensions (34, 48) and are called the skeleton notes. The skeleton note attracts the audience’s attention and makes a deeper impression on them. By contrast, the remaining part of the notes plays a decorative role, giving the melody personalities or styles, and are called the decorative notes or prolongation notes. The melodic skeleton serves as the crucial structural support of rhythm and harmony, indicating the direction of melody development. Knowledge of melodic skeleton information can help humans and machines better analyze, understand, and learn the logic of melodic hierarchy organization from surface to deep layers.

Here, we introduce a framework for melodic skeleton extraction based on the knowledge of music theory (34, 35, 50–53) and music psychology (48, 49) to identify the dominant and subordinate relationship of structural importance from rhythm and pitch dimensions between melody notes. The theoretical basis and implementation details are described briefly below.

\subsubsection{Rhythmic skeleton extraction}
Before describing the theoretical basis of this work, we would have to cover some basic musical terms and concepts regarding meter and rhythm in the time aspect of music. In music theory, the pulse splits time into a series of uniformly spaced chunks called beats. Not all beats are created equal, and certain beats are felt stronger than others. The first beat of each measure is a downbeat, and the one that follows is an upbeat. The meter measures the number of beats in the regular and repeated pattern of the downbeat. For example, the most common meter in music is 4/4; each measure has four quarter note beats. The distribution pattern of strong and weak beats is “strong weak sub-strong weak,” as shown in Fig. 2A. Rhythm can be defined as the organization pattern of one downbeat with one or more upbeat (54). Therefore, the meter provides the temporal framework for organizing music rhythm.  

To draw listeners’ attention, musicians often use accents in musical compositions or performances to emphasize a particular note. Accents can be expressed in various ways to increase musical expressiveness and add character to the movement of music. Metrical and rhythmic accents are the two main accents in the symbolic melody data. A metrical accent is an accent that falls on the strong beat position within a measure. The metrical accent is periodic and cyclic, whose distribution pattern depends on the type of meter. A rhythmic accent is an accent in a strong position within the rhythm, which emphasizes a point that is not constrained by the meter’s structure. Consequently, it is flexible and changeable rather than static and fixed. Rhythmic accents can be created by increasing the notes’ dynamic (i.e., dynamic accent), extending the notes’ duration (agogic accent), using syncopation, and so forth. The agogic accent can be easily distinguished by comparing the surrounding notes on duration. Additionally, the musician can use syncopation to change the normal rhythm pattern by extending the duration of notes from the weak beat or weak beat part to the subsequent strong beat or strong beat part. Note that a note cannot be both a metrical accent and a syncopation, which are conflicted with each other. The dynamic accent is not used since it has no obvious changes in most symbolic melody data.  

In this study, we extract the metrical accents, the agogic accents falling on the metrical accent, and the agogic accents falling on the syncopation as the rhythmic skeleton notes, as illustrated in Fig. 2B. Metrical accents are the foundation of other types of accent (34). When two or more different types of accents work together, the listener will experience a particularly more prominent accent. Therefore, when a rhythmic accent and a metrical accent overlap, the rhythmic accent is generally more perceptible and is the one that is preferred. If there were continuous rhythmic skeleton notes, we chose the most structurally important note as the rhythmic skeleton note according to the intensity of the accent.

\subsubsection{Tonal skeleton extraction}
In tonal music, the pitch of the melody moves around a centrally stable note (i.e., the tonal center or tonic), repeatedly moving away from it and back to it. Pitches are essentially organized into a distinct hierarchy scale based on tonal stability. There is a mutual attraction between pitches with different stable levels, which can stimulate different emotional experiences (47). Specifically, an unstable pitch tends to be a stable pitch, which would make the listeners feel relaxed or dismissed. Moving from a stable to an unstable pitch would increase the listener’s sense of tension. The prolongational reduction theory of GTTM suggests that the more important music event has less tension and vice versa (53). Note that the same tone may have different feelings in different contexts, which may be pleasant or anxious.  

For the tonal skeleton extraction method, we use the tension level as a metric to quantify the relative importance of the pitch. The specific recognition procedure is as follows.  
\begin{itemize}
\item First, we used the position of the rhythmic skeleton notes as the boundary of the individual context because the metrical structure is the important basis of all hierarchical structure types (34).  
\item Second, we combined two or three successive notes as the minimum rhythmic cell in each segment, according to the repetition frequency in the melody (55) and the number of notes of this rhythmic cell. The term “rhythmic cell” defines as a “small rhythmic and melodic design that can be isolated or can make up one part of a thematic context” (56). Therefore, each rhythmic cell can be seen as an isolated thematic context for calculating the tension proﬁle.  
\item Finally, we adopted a mathematical tonal tension model to quantify each note’s tension value by calculating the distance between every single tone and global key in the spiral array (57–59), as shown in Fig. 2C. We selected the note with the minimum tension value in each rhythmic cell as the tonal skeleton note. For example, Fig. 2D shows the tonal skeleton of the first eight bars from the song “Hey Jude.”  
\end{itemize}

\subsection{Design of WuYun}
Figure 1C shows the diagram of the proposed hierarchical melody generation architecture called WuYun. First, we convert the melody MIDI files and their melodic skeletons into musical event sequences as the input data for model training using the MeMIDI symbolic music representation method. Then, we design a hierarchical melody generation architecture with two generative modules responsible for melodic skeleton construction and melody inpainting, respectively. In this subsection, we will introduce the hierarchical melody generation architecture about how we generate the melodic skeleton and incorporate it to guide the melody generation process. The details about the MeMIDI symbolic music representation method and the word embedding technique used in this architecture’s input module are described in the Materials and Methods section.  

WuYun is designed to generate melodies in two stages hierarchically: melodic skeleton construction and melody inpainting, instead of the dominant end-to-end left-to-right note-by-note melody generation paradigm. At the stage of melodic skeleton construction, we use the Transformer-XL model with only the decoder as the melodic skeleton generation module (19), which has the advantage of remarkable performance in capturing long-term dependence. To develop the capacity of melodic skeleton construction, we trained the Transformer-XL model on the extracted melodic skeleton database. At the stage of melody inpainting, we employ the recurrent Transformer-based encoder–decoder architecture (18) in a sequence-to-sequence setup as the melody inpainting module to complete the melody conditioned on the melodic skeleton, i.e., filling the missing information between the melodic skeleton notes. In this work, the melody inpainting problem can be defined as follows: given a melodic skeleton sequence $C_{s}$, generate an inpainted melody sequence $C_{m}$. The encoder maps the discrete input symbols of the melodic skeleton sequence $C_{s}$ to a high-dimensional continuous vector as conditional input into the decoder, and the decoder then generates an output sequence $C_{m}$ in an autoregressive manner. The melodic skeleton sequence will be saved in the final generated melody. This method provides users an entry point to interact with the melody generation model by adjusting melodic skeleton notes between two stages to control the melodic motion.  

In this work, we focus on designing a hierarchical skeleton-guided melody generation architecture based on knowledge-enhanced deep learning, following the hierarchical organization principle of structure and prolongation. Therefore, we used common language models in NLP to make the WuYun architecture accessible. The capacity of these two generative modules may be further optimized; however, the goal of this study is not to find the most optimal neural network.

\subsection{Evaluation metrics}
Subjective and objective evaluations are the two essential aspects of evaluating the performance of music generation systems. The human listening test is currently an indispensable and viable method for subjective evaluation to measure the quality of the generated musical pieces. However, for the objective evaluation, many eﬀorts have been made to design quantitative metrics; there is not a set of convincing and unified metrics. Although the research field of music generation is multidisciplinary, most researchers mainly focus on generative models with different improvement goals rather than their contribution to quantifying music complexity. Consequently, almost all the proposed objective evaluation metrics are difficult to apply for comparing different music creation systems and lack sustainability for future development demands. We tried to calculate the averaging overlapped area of some musical feature distributions between generated musical pieces and ground-truth musical pieces as the objective evaluation metrics like (18, 60) using the public evaluation toolbox. We arrived at a similar conclusion as PopMNet (32) that a better result of objective evaluation does not mean better structure and musicality of generated music. The same objective evaluation result can be calculated and verified with the provided melody MIDI files of this study’s next two experiments. Therefore, we conducted two subjective evaluation experiments to evaluate the performance of our proposed WuYun, including different melodic skeleton settings in rhythm and pitch dimensions and comparisons with public state-of-the-art (SOTA) music generation models.  

We randomly selected ten melodies from the evaluation dataset for the listening materials. Similar to previous studies (15, 17, 19), we took the first four bars as prompt and set the maximum number of generated bars to 28. We assigned a random order for all musical pieces as the file name, including the generated and ground-truth musical pieces. All melody MIDI files were rendered into audio via a piano MIDI synthesizer. In the blind listening test, participants were asked to rate each melody on a ﬁve-point Likert scale (i.e., 1 for bad and 5 for excellent) on five dimensions:  
\begin{itemize}
\item \textbf{Rhythm:} Whether the brain can feel the regular accents and rhythm patterns.
\item \textbf{Richness:} Whether the melody sounds rich and interesting in the rhythm and pitch dimensions.
\item \textbf{Structure:} Whether the brain can feel the boundary of melodic phrases and the balance among melodic phrases’ length.
\item \textbf{Expectation:} Whether the direction of melody development meets the audience’s expectations for melody development (61).
\item \textbf{Overall:} Overall quality.
\end{itemize}
We found that most nonmusicians had heard technical music terms but did not understand what they meant. Therefore, before formal experiments, we conducted multiple rounds of discussions, testing, and validation with musicians and nonmusicians regarding the above subjective evaluation metrics and their descriptions until they could easily understand and grasp them. To ensure that the recruited subjects have a common understanding of the metrics and scales in the questionnaire, we also conducted evaluation training for them, including the explanation of subjective evaluation metrics and preliminary experiments. All audio and MIDI files for evaluation can be found in Acknowledgments.

\subsection{Model performance based on different melodic skeleton settings}
To compare the effectiveness of variants of melodic skeleton extracted from rhythm and pitch dimensions, we comprehensively evaluated the performance of WuYun based on different settings of the melodic skeleton. Furthermore, we added three control group settings of randomly selected notes with different percentages as the melodic skeleton in order to verify the effectiveness of the proposed melodic skeleton extraction method based on music domain knowledge. All experimental settings and the proportion of melodic skeleton notes in the melody are described below: 
\begin{enumerate}
    \item \textbf{Downbeat} only uses metrical accents as the melodic skeleton (32.8\%).  
    \item \textbf{Long Note} only uses agogic accents as the melodic skeleton (27.4\%).  
    \item \textbf{Rhythm} uses rhythmic skeleton notes as the melodic skeleton (33.8\%).  
    \item \textbf{Tonic} uses tonal skeleton notes as the melodic skeleton (43.2\%).  
    \item \textbf{Interaction} uses the intersection of rhythmic skeleton notes and tonal skeleton notes as the melodic skeleton (14.2\%).  
    \item \textbf{Union} uses the union of rhythmic skeleton notes and tonal skeleton notes as the melodic skeleton (62.8\%).  
    \item \textbf{Random25\%} randomly selects 25\% of melody notes as the melodic skeleton (25\%).  
    \item \textbf{Random50\%} randomly selects 50\% of melody notes as the melodic skeleton (50\%).  
    \item \textbf{Random75\%} randomly selects 75\% of melody notes as the melodic skeleton (75\%).  
\end{enumerate}
In this experiment, we obtained 90 musical pieces for rating. We recruited 30 subjects (13 females and 17 males, ages 18 and 30 years) from Zhejiang University and Zhejiang Conservatory of Music to evaluate the musical pieces with payment. Fifteen subjects among them were professional music practitioners with an average of 9 years of music training and 4 years of music performance experience. The rest of the subjects have little professional music training or performance experience. Each musical piece was assigned to three professionals and three nonprofessional subjects. Each subject was required to rate 18 musical pieces, which cost approximately 25 min.  

Figure 3A shows the mean opinion scores of WuYun architecture's melody generation performances with nine different settings on the five subjective evaluation metrics from all subjects in the form of histograms. The detailed experimental result is shown in Table S1. Generally, among all melodic skeleton settings, the proposed rhythmic and tonal skeleton based on music theory and psychological study performs better than other skeletons. The rhythmic skeleton setting (No. 3) achieved the best result on all subjective evaluation metrics, followed by the tonal skeleton setting (No. 4). Among the three types of melodic skeletons associated with rhythm (Nos. 1, 2, and 3), the melodic skeleton composed of a single type of accent (e.g., metrical accents or agogic accents (31)) has a large gap with the rhythmic skeleton in richness, expectation, and overall quality and even surpassed by the random melodic skeletons (Nos. 7, 8, and 9) on most subjective evaluation metrics. This result indicates that a flexible rhythmic skeleton (i.e., including several kinds of musical accents) is essential for melody composition to improve musicality. In contrast, a rigid melodic skeleton (i.e., including only one type of accent, especially metrical accents) reduces the quality of the generated melody and limits the models' performance. Likewise, as depicted in the right part of Fig. 2C, the pitch classes of the tonal skeleton notes are mostly C, D, and E, which also lead to the rigidity of the generated tonal skeletons.  

Additionally, compared to the rhythmic and tonal skeleton settings, the intersection (No.5) and union (No. 6) skeleton settings led to a distinct degradation of the melody generation performance. For instance, the intersection (No. 5) skeleton setting received the worst scores in most evaluation aspects, even worse than the random sampling skeleton settings. This phenomenon can be explained by the structure and prolongation proportion tradeoffs in the design of two-stage melody generation architecture using an end-to-end learning framework. We can preliminarily see that with the increased percentage of melodic skeleton notes, the performance of the two-stage melody generation went up first but then down. On the one hand, a low proportion of melodic skeleton notes makes it easier to train the melodic skeleton construction model in the first stage. However, the generated skeleton notes will be too sparse to guide the second stage of melodic inpainting (such as the intersection skeleton setting, only 14.2\%). On the other hand, if the proportion of melodic skeleton notes is too large, the training difficulty and data dependency of the melodic skeleton construction model will increase. Besides, from the perspective of the gestalt theory (62) about the law of the figure–ground relationship, during the perception of music, a person’s attention constantly switches between different musical elements; sometimes, he/she may be attracted to the rhythm, whereas at other times, to the pitch. Therefore, the extraction of the hierarchical dependency structural relationship between musical elements is affected by multiple musical dimensions; it will not be like a simple addition or subtraction operation but a complex organic combination (63).  

In this study, we chose the setting of the rhythmic skeleton (No. 3) that performed best on all subjective evaluation metrics in this experiment as the default skeleton configuration (denoted as WuYun-RS) for the next experiment to compare with other melody generation models.

\begin{figure}
\centering
\includegraphics[width=1\textwidth]{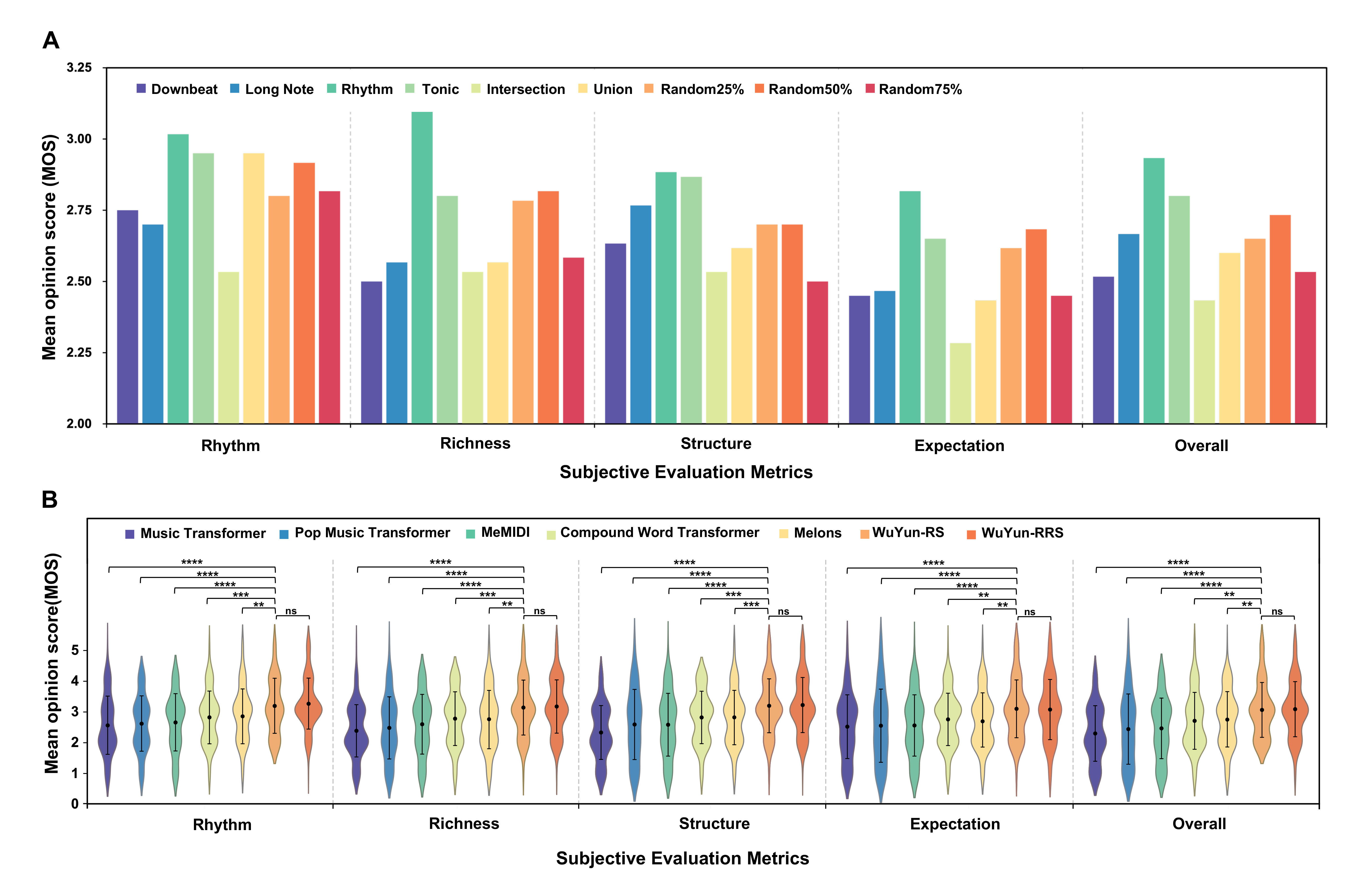}
\caption{\textbf{Subjective evaluation results of the WuYun melody generation architecture based on different melodic skeleton settings, and the other public melody generation models.} (\textbf{A}) Subjective comparison of the performance of the WuYun architecture based on different melodic skeleton settings in Experiment 1. Data is the mean opinion score. The WuYun architecture with the rhythmic skeleton setting achieves the best performance in all melodic skeleton settings on all subjective evaluation metrics. (\textbf{B}) Subjective comparison of the performance of different music generation models in Experiment 2. Violin plots show the kernel density estimate of rating distribution, the larger the area of the area graph, the greater the probability of the value distribution. The black dot within the violin plots indicates the mean; the black line within the violin plot indicates the standard deviation. Statistical analyses are done between WuYun-RS and the rest of the models using the one-tailed t-test (N = 130). P values of statistical significance are represented as *$P < 0.05$, **$P < 0.01$, ***$P < 0.001$, and ****$P < 0.0001$; ns, insignificant. The subjective comparison results' data are in Tables S1, 1, and S2, respectively.
}
\label{fig_3}
\end{figure}

\subsection{Comparisons with other melody generation methods}
To prove the effectiveness of the proposed hierarchical skeleton-guided melody generation architecture based on knowledge-enhanced deep learning, we compared WuYun-RS (i.e., using the rhythmic skeleton setting) to five public SOTA Transformer-based melody generation models, namely, Music Transformer (15), Pop Music Transformer (17), Compound Word Transformer (19), Melons (33) and MeMIDI, that follow an end-to-end left-to-right note-by-note generative paradigm and treat each note equally. The MeMIDI setting uses the MeMIDI data representation method like WuYun-RS and employs the Transformer-XL model without using the melodic skeleton for the melody generation task. Moreover, to prove the effectiveness of the generated melodic skeleton, we added the setting of WuYun-RRS, skipped the melodic skeleton construction in the first stage, and directly used the real rhythmic skeleton as additional knowledge to guide the melody generation process of melody inpainting in the second stage. However, the original music representation of Music Transformer does not include chord progressions. For a fair comparison, we added the CHORD events proposed in this work into the MIDI-Like music representation of Music Transformer. Each CHORD event was followed by a TIME-SHIFT event and had a higher sorting priority than NOTE-related events. Additionally, the MIDI quantization level of the Pop MusicTransformer, Compound Word Transformer, and Melons only considered the 16th note time grid. Therefore, in this experiment, we applied the 16th note time grid as the MIDI quantization level to the melody dataset for all music generation models.  

In this experiment, we obtained 80 musical pieces for rating. Since the second subjective evaluation experiment relies on the result of the first subjective evaluation experiment and requires some time to collect, process, and analyze, we recruited 13 participants again (i.e., six females and seven males, ages 18 and 25 years) to evaluate the musical pieces with payment. Six subjects among them were professional music practitioners with an average of 8 years of music training and 4 years of music performance experience. Each subject was required to rate all musical pieces. After rating 20 musical pieces, subjects were asked to rest for 5 min against hearing fatigue. The average experiment time cost each subject about 2 h.  

Figure 3B shows the mean opinion scores and one-tailed t-test results of the different music generation systems on the five evaluation metrics in the form of violin plots. The detailed experimental results are shown in Tables 1 and S2. Overall, WuYun-RS (No. 6) and WuYun-RRS (No. 7) outperformed the other five current SOTA end-to-end left-to-right note-by-note melody generation models on all metrics, including MusicTransformer (No. 1), Pop Music Transformer (No. 2), Compound Word Transformer (No. 4), Melons (No. 5), and MeMIDI (No. 3). Besides, except for WuYun-RRS, there is a significant difference ($P < 0.01$) between WuYun-RS and the other melody generation systems. This result demonstrates that WuYun-RS and WuYun-RRS are able to generate melodies with improved long-term structure and musicality, which benefit from the rhythmic skeleton as a deep structure to guide the melody generation process. Furthermore, WuYun-RS and WuYun-RRS demonstrate highly similar performances in terms of the quality of generated melodies on all evaluation metrics. This result indicates the effectiveness of the melodic skeletons generated via the melodic skeleton construction module. However, there is still an obvious gap between the WuYun melody generation architecture and human-composed music, leaving room for improvement. This also shows that designing clever decorations for melodic skeletons is another difficult research problem, even for human composers. Additionally, when using the same symbolic music representation method, the knowledge-enhanced hierarchical skeleton-guided melody generation model of WuYun-RS greatly outperformed the single-stage end-to-end left-to-right note-by-note melody generation model of MeMIDI (No. 3). On the one hand, this demonstrates that our proposed hierarchical melody generation paradigm can be applied to empower the dominant end-to-end left-to-right note-by-note melody generation paradigm. On the other hand, although the Compound Word Transformer and Melons (Nos. 4 and 5) were inferior to WuYun-RS, their effective compound word representation and the linear Transformer as the backbone architecture enable it to process multidimensional music information in one step simultaneously and obtain a better result among these five public SOTA melody generation models. Thus, combining the proposed knowledge-enhanced hierarchical skeleton-guided music generation architecture with more efficient music representation methods and advanced language models can bring a better result for melody generation tasks.

\begin{table}
  \begin{center}
\renewcommand\arraystretch{1.25}
  \caption{Subjective evaluation scores of generated melodies based on different melody generation models in Experiment 2 (mean $\pm$ standard deviation).}
  \label{sample-table}
\begin{tabular}{
>{\columncolor[HTML]{ffffff}}c 
>{\columncolor[HTML]{ffffff}}c 
>{\columncolor[HTML]{ffffff}}c 
>{\columncolor[HTML]{ffffff}}c 
>{\columncolor[HTML]{ffffff}}c 
>{\columncolor[HTML]{ffffff}}c 
>{\columncolor[HTML]{ffffff}}c }
\hline
\textbf{No.} & \textbf{Model} & \textbf{Rhythm} & \textbf{Richness} & \textbf{Structure} & \textbf{Expectation} & \textbf{Overall} \\ \hline
1            & MT             & 2.52 $\pm$ 0.93       & 2.34 $\pm$ 0.83    & 2.28 $\pm$ 0.86          & 2.47 $\pm$ 1.01            & 2.25 $\pm$ 0.88        \\
2            & PMT            & 2.57 $\pm$ 0.88       & 2.43 $\pm$ 0.99    & 2.54 $\pm$ 1.11          & 2.50 $\pm$ 1.16            & 2.39 $\pm$ 1.12        \\
3            & MeMIDI         & 2.61 $\pm$ 0.91       & 2.55 $\pm$ 0.95    & 2.53 $\pm$ 1.00          & 2.51 $\pm$ 0.97            & 2.42 $\pm$ 0.96        \\
4            & CWT            & 2.77 $\pm$ 0.83       & 2.72 $\pm$ 0.85    & 2.74 $\pm$ 0.81          & 2.70 $\pm$ 0.83            & 2.65 $\pm$ 0.90        \\
5            & Melons         & 2.84 $\pm$ 0.89       & 2.68 $\pm$ 0.95    & 2.75 $\pm$ 0.87          & 2.67 $\pm$ 0.87            & 2.71 $\pm$ 0.88        \\
6            & WuYun-RS       & 3.13 $\pm$ 0.88       & 3.07 $\pm$ 0.87    & 3.13 $\pm$ 0.86          & 3.02 $\pm$ 0.92            & 3.00 $\pm$ 0.87        \\
7            & WuYun-RRS      & 3.20 $\pm$ 0.81       & 3.11 $\pm$ 0.85    & 3.15 $\pm$ 0.88          & 3.00 $\pm$ 0.96            & 3.02 $\pm$ 0.88        \\
8            & Human          & 3.54 $\pm$ 0.82       & 3.65 $\pm$ 0.76    & 3.68 $\pm$ 0.89          & 3.55 $\pm$ 0.92            & 3.57 $\pm$ 0.84        \\ \hline
\end{tabular}
  {\small
  \begin{tablenotes}
     \item[1]MT, PMT, and CWT stand for Music Transformer, Pop Music Transformer, and Compound Word Transformer, respectively.
   \end{tablenotes}
    }
\end{center}
\end{table}

\section{Discussion}
The methodology we have taken in designing WuYun, a hierarchical skeleton-guided melody generation architecture based on knowledge-enhanced deep learning, combines music analysis theory and musical psychology. Unlike the dominant end-to-end left-to-right note-by-note melody generation paradigm, we use the hierarchical organization principle of structure and prolongation to decompose the melody generation process into melodic skeleton construction and melody inpainting stages. We extract the most structurally important notes based on hearing sensitivity as melodic skeletons and incorporate them into the melody generation process as a deep structure to guide the model to learn the hierarchical dependency structures among musical event sequences from the limited melody data without music boundary detection (31). The human evaluation results demonstrated that our model exhibits significant improvement in both long-term structure and musicality across the structured melody generation task.  

In practical application scenarios, the ability to obtain real feedback from human users for improving the performance and interaction experience of the system is essential for the next generation of iterative and interactive music generation systems. In general, WuYun allows human users to edit the generated melodic skeleton and adjust its shape to guide and constrain the range of the decorative notes at the next stage of melody inpainting. Thus, the proposed generation strategy based on the hierarchical organization principle of structure and prolongation not only can maintain the long-range tonal coherence of generated melodies but also achieve control over the target of melodic motion by human users. Additionally, with WuYun and its melodic skeleton analysis framework, human users can directly extract the skeleton from existing music compositions for music composition analysis or re-creation.  

Our study has some limitations, notably the performance of the melody inpainting model, except that the quality of generated melodic skeleton may be poor; even if an original rhythmic skeleton is provided, the quality of the completed melody is still far from the real one. Further performance improvements could be achieved using pretrained masked language models (23) for music generation, especially for the melody inpainting task. Another issue is how to effectively extract an organic melodic skeleton from hierarchical musical structures combining two or more musical dimensions (e.g., rhythm and pitch) to further improve the structure of generated melodies. According to the research in the cognitive psychology in music, while listening to music, only by combining the tonal and rhythmic structures can we form a more coherent musical representation and create a complete sense of melody (34, 48, 49). However, the brain's processing mechanism of the hierarchical musical structure remains a fundamental research problem in the field of music cognitive psychology (64–66). With the help of advanced electroencephalography devices, cognitive musicology can break through the human cognition of hierarchical musical structure and apply it to music generation. Additionally, we expect to investigate other systematic music analysis theories and gain further psychological knowledge to analyze and compare the hierarchical levels of important musical events along different musical dimensions for designing a more effective melodic skeleton extraction framework. Another direction is to explore explainable AI for music generation to assist end users in making better decisions since deep learning methods lack physical transparency of methods. With these potential future improvements in mind, we hope that our findings for structured melody generation will optimize the dominant melody generation paradigms to improve long-term structure and musicality and provide a new lens to develop multidisciplinary research via combining data-driven and knowledge-based approaches.

\section{Materials and Methods}

\subsection{Details of dataset preprocessing}
We evaluate the effectiveness of WuYun architecture on a commonly used and publicly available symbolic melody dataset of Wikifonia (32, 33, 67). The Wikifonia dataset contains thousands of lead sheets in MusicXML format. It covers various music genres, composed of melody and the accompanying chord progression and tonality labels. Here, we describe the procedure below to clean up noisy data and artificial errors since the dataset is user-generated.  
\begin{itemize}
    \item \textbf{Data Segmentation:} To simplify rhythm modeling, we only keep those segments from MIDI files with the most commonly used 4/4 time signature (18).
    \item \textbf{MIDI Quantization:} For a more beat-accurate timing of sounds, quantization is a useful digital music processing of setting MIDI data on beats or exact fractions of beats to eliminate some imprecise timing because of expressive musical performance or mistake record. We contend that a more precise and adaptable time grid is required to model a more expressive metrical context, including the 32nd (18), 64th note, and even triplets. By contrast, most prior works only use the 16-note time grid for quantification (17, 19, 31–33); each bar is quantized into 16 intervals. In this study, we propose a self-adaptive mixed precision quantization method to reduce quantization errors (Fig. 4). This method can automatically choose a suitable quantize grid for every single note based on its duration, including straight notes and triplets. The difference between straight notes and triplets is dividing the musical beat evenly in half or third. First, the notes shorter than a 64th note are discarded, whereas notes longer than one bar are saved into the whole note. Second, according to the note duration, the rest notes are classified into straight or triplets. However, triplets do not always have to have three notes. There are only two notes in triplets is quite common, such as in Swing. Additionally, in theory, every note in triplets has an equal rhythm value. However, in practice, most notes are slightly different from each other in musical performance. Therefore, two or three consecutive notes with approximately the same duration and consistent with the duration of triplets are considered as triplets. Based on our experimental statistical results, we set the acceptable duration error ratio between the actual triplet and the standard triplet to within 20\%. Last, the method automatically selects a proper quantize grid for every note. In terms of straight notes, the granularity of the time grid depends on the note duration. Particularly, the note onset is aligned to its closest 16th note time grid when the note duration is greater than or equal to a 16th note, to its closest 32nd note time grid when the note duration is between a 16nd and 32th note, and to its closest 64th note time grid when the note duration is between a 32nd and 64th note. Moreover, the straight note offset is aligned to the 64th note time grid. In terms of triplets, the note onset and offset time is aligned to the 48th note time grid.
    \item \textbf{Tonality Unification:} For simplicity, the tonalities and chord progressions of those MIDI files are transposed to “C major” and “A minor” tonalities (68). We set one chord per beat and unify the chord representation of the Wikifonia dataset using the chord dictionary as described in the following subsection.
    \item \textbf{Octave Transposition:} All melodies are applied octave transposition to shift the pitch into the range from C3 to C5 or are removed, which are out of the regular melodic pitch range (32).
\end{itemize}
After data cleaning, we get 2,921 musical pieces in Wikifonia, including 116,935 bars and 425,223 notes. Finally, randomly hold out 50 songs for testing and use the remaining for training.

\begin{figure}
\centering
\includegraphics[width=1\textwidth]{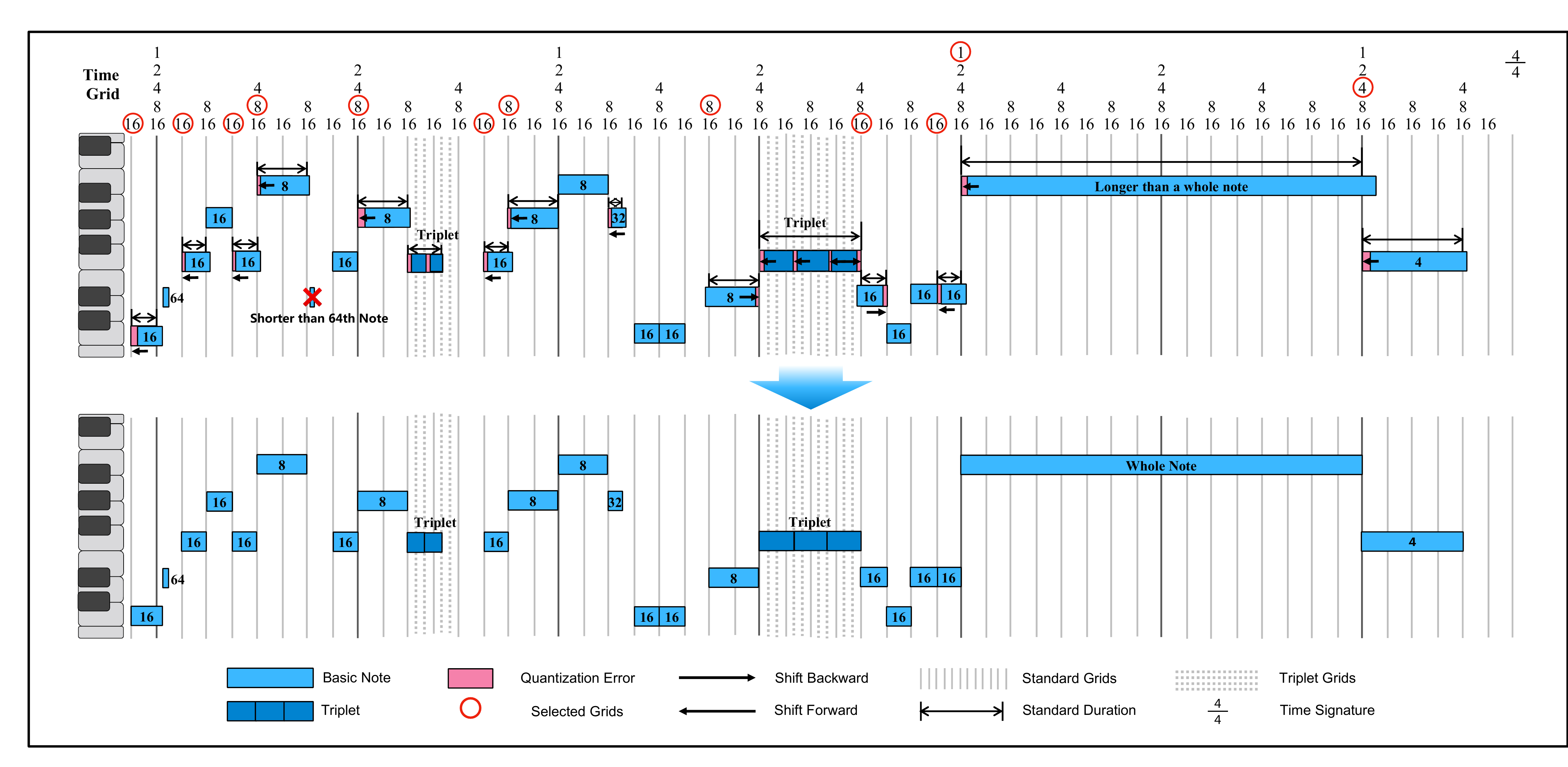}
\caption{\textbf{ Illustration of the self-adaptive mixed precision MIDI quantization.} The MIDI quantization method would automatically choose a suitable quantize grid for every note according to its note length to eliminate imprecise timing, including straight notes (minimum 64th note) and triplets (minimum 48th note).
}
\label{fig_4}
\end{figure}

\subsection{Symbolic melody representation}
In this work, we adopted a modiﬁed version of the “MuMIDI” symbolic music representation (18) to encode a piece of monophonic melody into discrete musical event sequences. We refer to it as MeMIDI. Following is a description of the MeMIDI extensive symbols information, which includes bar, position, note, chord, and tempo symbols.
\begin{itemize}
    \item \textbf{Bar and Position} \\ We use a bar symbol to represent a bar line and a position symbol to represent the onset of a note or a chord event. Since the minimum time grids of the straight and triplet note are 64th and 48th notes, respectively, and the MIDI files’ time resolution is 480 ticks per beat; thus we merge these two kinds of minimum time grids values ($\left \{ 0,30,60,...,1890 \right \} \cup \left \{  0,40,80,...,1880\right \} $) and use the <\textit{Pos\_Value}> symbol to represent 96 kinds of starting positions, such as <\textit{Pos\_30}>. We assign a position symbol for every chord and note music event.
    \item \textbf{Note} \\ A note has three basic attributes: pitch, duration, and velocity. Here, the value of the note pitch attribute ranges from 48 (C3) to 83 (C5). The value of the note velocity attribute ranges from 0 to 127. Considering both straight notes and triplet notes, the value range of the note duration attribute is $\left \{ 30, 60, 90,...,1920 \right \} \cup \left \{ 40,80,160,320,640 \right \} $ ticks. We use a compound word <\textit{Pitch\_Value, Velocity\_Value, Duration\_Value}> to compress these three attributes of one note in one token to shorten the length of the melody events sequence.
    \item \textbf{Chord} \\ To cover the chord types in the Wikifonia dataset, we use a more comprehensive chord event list. As shown in Table 2, we consider 12 chord roots and 13 chord qualities, yielding 156 possible chord events. We use a chord symbol <\textit{Root\_Quality}> to represent a chord musical event. To reduce repetition, we use the same position symbol for Note and Chord, which start at the same time. For simplicity, we do not use the Chord symbol in the melodic skeleton event sequence.
    \item \textbf{Tempo} \\ We divide the tempo into three categories: low (below 90), medium (90 to 160), and high (above 160).
        
\end{itemize}

\begin{table}
\renewcommand\arraystretch{1.25}
  \caption{List of chord events.}
  \begin{center}
  \label{table_chord}

\begin{tabular}{|
>{\columncolor[HTML]{FFFFFF}}c |
>{\columncolor[HTML]{FFFFFF}}l 
>{\columncolor[HTML]{FFFFFF}}l |}
\hline
\textbf{Chord}                                          & \multicolumn{2}{l|}{\cellcolor[HTML]{FFFFFF}\textbf{Content}}                                     \\ \hline
Chord root                                              & \multicolumn{2}{l|}{\cellcolor[HTML]{FFFFFF}C, Db, D, Eb, E, F, F\#, G, Ab, A, Bb, B}             \\ \hline
\cellcolor[HTML]{FFFFFF}                                & \multicolumn{1}{l|}{\cellcolor[HTML]{FFFFFF}Triad}         & M, m, o, +                           \\ \cline{2-3} 
\cellcolor[HTML]{FFFFFF}                                & \multicolumn{1}{l|}{\cellcolor[HTML]{FFFFFF}Seventh chord} & MM7, Mm7, mM7, mm7, o7, \%7, +7, +M7 \\ \cline{2-3} 
\multirow{-3}{*}{\cellcolor[HTML]{FFFFFF}Chord quality} & \multicolumn{1}{l|}{\cellcolor[HTML]{FFFFFF}Suspension}    & Sus.                                 \\ \hline
\end{tabular}
\end{center}
\end{table}

\subsection{WuYun architecture}
Here, we brieﬂy elaborate on the configuration details of the two Transformer-based generative modules of WuYun architecture, i.e., the melodic skeleton generation module for the melodic skeleton construction stage and the melodic prolongation generation module for the melody inpainting stage. We refer readers to (17–19) for more details. For reproducibility, we do not tweak the architecture of referenced models so that our music generation architecture can be easily assembled with the public implementation of Transformers.  

We use an unconditional sequence learning model Transformer-XL for the melodic skeleton generation module. We use four self-attention layers, each with eight attention heads. The model hidden size and the inner layer of the feed-forward part are set to 512 and 2,048, respectively. All token embedding sizes are set to 512, following (19). We use the compound word embedding (Fig. S1) and token attribute prediction method for the input and output modules, repectively (18). We employed the top-k temperature-controlled stochastic sampling method ($k = 10$, $temperature = 0.9$) during inference. The length of training input tokens and the memory length are also 512. Here, we used the melodic skeleton data extracted from the training part of the Wikifonia dataset to train the melodic skeleton generation module.  

We use a conditional sequence-to-sequence model based on Transformer-based recurrent encoder–decoder neural networks for the melodic prolongation generation module (18). We set the number of encoder layers, decoder layers, encoder heads, and decoder heads to 4. The size of hidden layers and the dimension of token embeddings are set to 256. We keep the same input module, output module, sampling method, length of training input tokens, and memory as same as the melodic skeleton generation module. For training the melodic prolongation generation module, we use the MeMIDI representations of the paired melodic skeleton and melody data as the encoder and decoder input data, respectively.

\subsection{Training}
We implemented the WuYun architecture with Pytorch (v1.7.1) (69). The parameters of the WuYun architecture were optimized by minimizing the cross-entropy loss on a single NVIDIA GTX 2080-Ti GPU with 11 GB memory. Specifically, the training loss was minimized with the Adam optimizer ($\beta _{1}  = 0.9$, $\beta _{2}  = 0.98$), a learning rate of $\varepsilon =10^{-3} $ , and dropout was applied with a ratio of 0.1. The mini-batches of the input data for the melodic skeleton generation module and the melodic prolongation generation module were 20 and 44, respectively. It took nearly 2 days to train the two modules until training convergence.

\subsection{Statistical analysis}
All subjective evaluation results were expressed as mean $\pm$ standard deviation. The statistical significance of the performance difference in WuYun-RS and other melody generation methods was analyzed using the one-tailed t-test. Asterisk indicates significant difference at *$\textit{P} < 0.05$, **$\textit{P} < 0.01$, ***$\textit{P} < 0.001$, ****$\textit{P} < 0.0001$, and ns, not significant.

\section*{Acknowledgements}
Thanks to Huawei Technologies Co., Ltd for the help in dataset collection and comments. We thank Jiaxing Yu, Chongjun Zhong, Ruiyuan Tang, and Jiaqi Wang for insightful discussions and visualizations.

\textbf{Funding:} This work is supported by the National Natural Science Foundation of China (No.62272409), the Key R\&D Program of Zhejiang Province (No.2022C03126), the Project of Key Laboratory of Intelligent Processing Technology for Digital Music (Zhejiang Conservatory of Music), and the Ministry of Culture and Tourism (No.2022DMKLB001).

\textbf{Author contributions:} Conceptualization: K.Z., L.S. Methodology: X.W., T.Z., Z.H., K.Z. Investigation: T.Z., Z.H., Q.L. Visualization: S.W., Q.L. Supervision: L.S., K.Z., X.T. Writing—original draft: X.W., K.Z., Q.L. Writing—review \& editing: K.Z., X.T., X.W., L.S.

\textbf{Competing interests:}
K.Z., X.W., and T.Z. are inventors on a patent application related to this work filed by Zhejiang University. The authors declare that they have no other competing interests.

\textbf{Data and materials availability:}
All data needed to evaluate the conclusions in the paper are present in the paper and/or the Supplementary Materials. Raw experimental data and the generated symbolic melody files are available on Zenodo at DOI 10.5281/zenodo.7480957 under a Creative Commons Attribution 4.0 International license. The code of the WuYun music generation framework is available at \url{https://github.com/NEXTLab-ZJU/wuyun}.

\section*{References}
{
\small

[1] F. Carnovalini, A. Rodà, Computational creativity and music generation systems: An introduction to the state of the art. {\it Front. Artif. Intell. Appl.} \textbf{3}, 14 (2020).

[2] Y. LeCun, Y. Bengio, G. Hinton, Deep learning. \textit{Nature} \textbf{521}, 436–444 (2015).

[3] J. Briot, G. Hadjeres, F. Pachet, \textit{Deep Learning Techniques for Music Generation} (Springer, 2020).

[4] E. Frid, C. Gomes, Z. Jin, Music creation by example, in \textit{Proceedings of the 2020 CHI conference on human factors in computing systems} (CHI, 2020), pp. 1–13.

[5] C. Yu, Z. Qin, F. J. Martín-Martínez, M. Buehler, A Self-Consistent Sonification Method to Translate Amino Acid Sequences into Musical Compositions and Application in Protein Design Using Artificial Intelligence. \textit{ACS Nano} \textbf{13}, 7471–7482 (2019).

[6] S. Di, Z. Jiang, S. Liu, Z. Wang, L. Zhu, Z. He, Z. He, H. Liu, S. Yan, Video background music generation with controllable music transformer, in \textit{Proceedings of the 29th ACM International Conference on Multimedia} (MM, 2021), pp. 2037–2045.

[7] L.W. Wesseldijk, M. A. Mosing, F. Ullén,. Why Is an Early Start of Training Related to Musical Skills in Adulthood? A Genetically Informative Study. \textit{Psychol. Sci.} \textbf{32}, 3–13 (2020).

[8] W. Zhou, C. Ye, H. Wang, Y. Mao, W. Zhang, A. Liu, C. Yang, T. Li, L. Hayashi, W. Zhao, L. Chen, Y. Liu, W. Tao, Z. Zhang, Sound induces analgesia through corticothalamic circuits. \textit{Science} \textbf{377}, 198–204 (2022).

[9] S. Koelsch, M. A. Rohrmeier, R. A. Torrecuso, S. Jentschke, Processing of hierarchical syntactic structure in music. \textit{Proc. Natl. Acad. Sci.} \textbf{110}, 15443–15448 (2013).

[10] A. D. Patel, Language, music, syntax and the brain. \textit{Nat. Neurosci.} \textbf{6}, 674–681 (2003).

[11] J. M. J. Valanarasu, P. Oza, I. Hacihaliloglu, V. M. Patel, Medical Transformer: Gated Axial-Attention for Medical Image Segmentation, in \textit{International Conference on Medical Image Computing and Computer-Assisted Intervention} (Cham, 2021), pp. 36–46.

[12] N. Li, S. Liu, Y. Liu, S. Zhao, M. Liu, M. T. Zhou, Neural speech synthesis with transformer network, in \textit{Proceedings of the AAAI Conference on Artificial Intelligence} (AAAI, 2019), pp. 6706–6713.

[13] P. Schwaller, B. Hoover, J. L. Reymond, H. Strobelt, T. Laino, Extraction of organic chemistry grammar from unsupervised learning of chemical reactions. \textit{Sci. Adv.} \textbf{7}, eabe4166 (2021).

[14] J. Jumper, R. Evans, A. Pritzel, T. Green, M. Figurnov, O. Ronneberger, K. Tunyasuvunakool, R. Bates, A. Žídek, A. Potapenko, A. Bridgland, C. Meyer, S. A. A. Kohl, A. J. Ballard, A. Cowie, B. Romera-Paredes, S. Nikolov, R. Jain, J. Adler, T. Back, S. Petersen, D. Reiman, E. Clancy, M. Zielinski, M. Steinegger, M. Pacholska, T. Berghammer, S. Bodenstein, D. Silver, O. Vinyals, A. W. Senior, K. Kavukcuoglu, P. Kohli, D. Hassabis, Highly accurate protein structure prediction with AlphaFold. \textit{Nature} \textbf{596}, 583–589 (2021).

[15] C. A. Huang, A. Vaswani, J. Uszkoreit, I. Simon, C. Hawthorne, N. Shazeer, A. M. Dai, M. D. Hoffman, M. Dinculescu, D. Eck, Music Transformer: Generating music with long-term structure, in \textit{7th International Conference on Learning Representations} (ICLR, 2018).

[16] B. Yu, P. Lu, R. Wang, W. Hu, X. Tan, W. Ye, S. Zhang, T. Qin, T. Liu, Museformer: Transformer with fine-and coarse-grained attention for music generation, in \textit{Advances in Neural Information Processing Systems} (NeurIPS, 2022).

[17] Y. Huang, Y. Yang, Pop music transformer: Beat-based modeling and generation of expressive pop piano compositions, in \textit{Proceedings of the 28th ACM International Conference on Multimedia} (MM, 2020), pp. 1180–1188.

[18] Y. Ren, J. He, X. Tan, T. Qin, Z. Zhao, T. Liu, Popmag: Pop music accompaniment generation, in \textit{Proceedings of the 28th ACM International Conference on Multimedia} (MM, 2020), pp. 1198–1206.

[19] W. Hsiao, J. Liu, Y. Yeh, Y. Yang, Compound word transformer: Learning to compose full-song music over dynamic directed hypergraphs, in \textit{Proceedings of the AAAI Conference on Artificial Intelligence} (AAAI, 2021), pp. 178–186.

[20] C. Payne. 2019. “Musenet.” OpenAI, July 21, 2022. http://openai.com/blog/musenet.

[21] N. Zhang, Learning adversarial transformer for symbolic music generation. \textit{IEEE Trans. Neural Netw. Learn. Syst.} 1–10 (2020).

[22] O. Sageev, S. Ian, D. Sander, E. Douglas, S. Karen, This time with feeling: Learning expressive musical performance. \textit{Neural. Comput. Appl.} \textbf{32}, 955–967 (2020).

[23] M. Zeng, X. Tan, R. Wang, Z. Ju, T. Qin, T. Liu, MusicBERT: Symbolic music understanding with large-scale pre-training, in \textit{Findings of the ACL: ACL-IJCNLP 2021} (ACL, 2021), pp. 791–800.

[24]  I. V. Anishchenko, T. M. Chidyausiku, S. Ovchinnikov, S. J. Pellock, D. Baker, De novo protein design by deep network hallucination. \textit{Nature} \textbf{600}, 547–552 (2020).

[25] R. Chowdhury, N. Bouatta, S. Biswas, C. Floristean, A. Kharkare, K. Roye, C. Rochereau, G. Ahdritz, J. Zhang, G. M. Church, P. K. Sorger, M. AlQuraishi, Single-sequence protein structure prediction using a language model and deep learning.\textit{ Nat. Biotechnol.} \textbf{40}, 1617–1623 (2022).

[26] J. Dauparas, I. V. Anishchenko, N. Bennett, H. Bai, R. J. Ragotte, L. F. Milles, B. I. M. Wicky, A. Courbet, R. Haas, N. Bethel, P. J. Y. Leung, T. F. Huddy, S. J. Pellock, D. Tischer, F. Chan, B. Koepnick, H. Nguyen, A. Kang, B. Sankaran, A. K. Bera, N. P. King, D. Baker, Robust deep learning-based protein sequence design using ProteinMPNN. \textit{Science} \textbf{378}, 49–56 (2022).

[27] J. Tang, X. Han, M. Tan, X. Tong, K. Jia, SkeletonNet: A Topology-Preserving Solution for Learning Mesh Reconstruction of Object Surfaces From RGB Images. \textit{IEEE Trans. Pattern Anal. Mach. Intell.} \textbf{44}, 6454–6471 (2020).

[28] D. Shi, X. Diao, H. Tang, X. Li, H. Xing, H. Xu, RCRN: Real-world Character Image Restoration Network via Skeleton Extraction. In \textit{Proceedings of the 30th ACM International Conference on Multimedia} (MM, 2022), pp. 1177–1185.

[29] K. Zhang, R. Zhang, Y. Yin, Y. Li, W. Wu, L. Sun, F. Wu, H. Deng, Y. Pan, Visual knowledge guided intelligent generation of Chinese seal carving. \textit{Front. Inf. Technol. Electron. Eng.} \textbf{23}, 1479–1493 (2022).

[30] J. Wu, C. Hu, Y. Wang, X. Hu, J. Zhu, A hierarchical recurrent neural network for symbolic melody generation. \textit{IEEE Trans. Cybern.} \textbf{50}, 2749–2757 (2019).

[31] S. Dai, Z. Jin, C. Gomes, R. B. Dannenberg, Controllable deep melody generation via hierarchical music structure representation, in \textit{Proceedings of the 22nd International Society for Music Information Retrieval Conference} (ISMIR, 2021), pp. 143–150.

[32] J. Wu, X. Liu, X. Hu, J. Zhu, Popmnet: Generating structured pop music melodies using neural networks. \textit{Artif. Intell.} \textbf{286}, 103303 (2020).

[33]Y. Zou, P. Zou, Y. Zhao, K. Zhang, R. Zhang, X. Wang, MELONS: generating melody with long-term structure using transformers and structure graph, in \textit{ICASSP 2022-2022 IEEE International Conference on Acoustics, Speech and Signal Processing} (ICASSP, 2022), pp. 191–195.

[34] F. Lerdahl, R. Jackendoff, \textit{A Generative Theory of Tonal Music} (MIT Press, Cambridge, MA, 1983).

[35] B. M. K. Ayotte, \textit{Heinrich Schenker: A Guide to Research} (Routledge, 2020).

[36] J. Berezovsky, The structure of musical harmony as an ordered phase of sound: A statistical mechanics approach to music theory. \textit{Sci. Adv.} \textbf{5}, eaav8490 (2019).

[37] F. Salzer, \textit{Structural Hearing: Tonal Coherence in Music} (Courier Corporation, 1962).

[38] Z. Dai, Z. L. Yang, Y. M. Yang, J. Carbonell, Q. Le, R. Salakhutdinov, Transformer-XL: Attentive language models beyond a fixed-length context, in \textit{Proceedings of the 57th Annual Meeting of the ACL} (ACL, 2019), pp. 2978–2988.

[39] A. Vaswani, N. Shazeer, N. Parmar, J. Uszkoreit, L. Jones, A. N. Gomez, L. Kaiser, I. Polosukhin, Attention is all you need, in \textit{Proceedings of the 31st Conference on Neural Information Processing Systems} (NIPS, 2017).

[40] G. Hadjeres, F. Nielsen, Anticipation-RNN: enforcing unary constraints in sequence generation, with application to interactive music generation. \textit{Neural. Comput. Appl.} \textbf{32}, 995–1005 (2020).

[41] B. Maess, S. Koelsch, T. C. Gunter, A. D. Friederici, Musical syntax is processed in Broca’s area: An MEG study. \textit{Nat. Neurosci.} \textbf{4}, 540–545 (2001).

[42] M. D. Hauser, N. Chomsky, W. T. Fitch, The faculty of language: What is it, who has it, and how did it evolve. \textit{Science} \textbf{298}, 1569–1579 (2002).

[43] W. T. Fitch, M. D. Hauser, Computational constraintson syntactic processing in a nonhuman primate. Science 303, 377–380 (2004).

[44]A. D. Friederici, J. Bahlmann, S. Heim, R. I. Schubotz, A. Anwander, The brain differentiates human and non-human grammars: Functional localization and structural connectivity. \textit{Proc. Natl. Acad. Sci.} \textbf{103}, 2458–2463 (2006).

[45] R. Näätänen, L. Anne, L. Mietta, C. Marie, H. Minna, I. Antti, V. Martti, Alku, Paavo, I. Risto, L. Aavo, Language-speciﬁc phoneme representations revealed by electric and magnetic brain responses. \textit{Nature} \textbf{385}, 432–434 (1997).

[46] A. Baddeley, Working memory: Looking back and looking forward. \textit{Nat. Rev. Neurosci} \textbf{4}, 829–839 (2003).

[47] H. Schenker, \textit{Neue Musikalische Theorien und Phantasien: Der Freie Satz} (Universal Edition, 1956).

[48] D. Hodges, \textit{Music in the Human Experience: An Introduction to Music Psychology} (Routledge, 2010).

[49] A. D. Patel, \textit{Music, Language, and the Brain} (Oxford University Press, 2010).

[50] D. Povel, Melody generator: a device for algorithmic music construction. \textit{J. Softw. Eng. Appl} \textbf{3}, 683 (2010).

[51] S. G. Laitz, \textit{The Complete Musician: An Integrated Approach to Tonal Theory, Analysis, and Listening} (Oxford University Press, 2012).

[52] F. Lerdahl, \textit{Tonal Pitch Space} (Oxford University Press, 2001).

[53] F. Lerdahl, C. Krumhansl, Modeling tonal tension. \textit{Music Percept.} \textbf{24}, 329–366 (2007).

[54] G. W. Cooper, L. B. Meyer, \textit{The Rhythmic Structure of Music} (Chicago University Press, 1963).

[55] C. L. Krumhansl, L. L. Cuddy, A theory of tonal hierarchies in music. \textit{Music Percept.} \textbf{36}, 51–87 (2010).

[56] Nattiez, J. J., \textit{Music and Discourse: Toward a Semiology of Music} (Princeton University Press, 1990).

[57] E. Chew, Mathematical and computational modeling of tonality, \textit{AMC} \textbf{10}, 141 (2014).

[58] D. Herremans, E. Chew, Tension ribbons: Quantifying and visualising tonal tension, in \textit{Proceedings of the Second International Conference on Technologies for Music Notation and Representation} (TENSOR, 2016), pp. 8–18.

[59] H. Dorien, E. Chew, MorpheuS: generating structured music with constrained patterns and tension. \textit{IEEE Trans. Affect. Comput.} \textbf{10}, 510–523 (2017).

[60] L. C. Yang, A. Lerch, On the evaluation of generative models in music. \textit{Neural. Comput. Appl.} \textbf{32}, 4773–4784 (2020).

[61] D. B. Huron, \textit{Sweet Anticipation: Music and the Psychology of Expectation} (MIT Press, Cambridge, MA, 2006).

[62] K. Koffka, \textit{Principles of Gestalt Psychology} (Routledge, 2013).

[63] A. M. Treisman, G. Gelade, A feature-integration theory of attention. \textit{Cogn. Psychol.} \textbf{12}, 97–136 (1980).

[64] R. M. Brown, J. L. Chen, A. Hollinger, V. B. Penhune, C. Palmer, R. J. Zatorre, Repetition suppression in auditory-motor regions to pitch and temporal structure in music. \textit{J. Cogn. Neurosci.} \textbf{25}, 313328 (2013).

[65] I. Peretz, M. Coltheart, Modularity of music processing. \textit{Nat. Neurosci.} \textbf{6}, 688–691 (2003).

[66] G. R. Kuperberg, T. F. Jaeger, What do we mean by prediction in language comprehension? \textit{Lang. Cogn. Neurosci.} \textbf{31}, 32–59 (2016).

[67] P. E. Hutchings, J. McCormack, Adaptive music composition for games. \textit{IEEE Trans. Games} \textbf{12}, 270–280 (2019).

[68] Z. Ju, P. Lu, X. Tan, R. Wang, C. Zhang, S. Wu, K. Zhang, X. Li, T. Qin, T. Liu, TeleMelody: Lyric-to-Melody generation with a template-based two-stage method, in \textit{Proceedings of the 2022 Conference on Empirical Methods in Natural Language Processing }(EMNLP, 2022), pp. 5426–5437.

[69] A. Paszke, S. Gross, F. Massa, A. Lerer, J. Bradbury, G. Chanan, T. Killeen, Z. Lin, N. Gimelshein, L. Antiga, A. Desmaison, A. Kopf, E. Yang, Z. De Vito, M. Raison, A. Tejani, S. Chilamkurthy, B. Steiner, L. Fang, J. Bai, S. Chintala, PyTorch: An imperative style, high-performance deep learning library. \textit{Adv. Neural Inf. Process. Syst.} \textbf{32}, 8026–8037 (2019).

}

\clearpage

\appendix

\section{Appendix}

\setcounter{table}{0} 
\setcounter{figure}{0} 
\renewcommand{\thetable}{S\arabic{table}}
\renewcommand{\thefigure}{S\arabic{figure}}

\subsection{Additional Supplementary Figure}
\begin{figure}[H]
\centering
\includegraphics[width=1\textwidth]{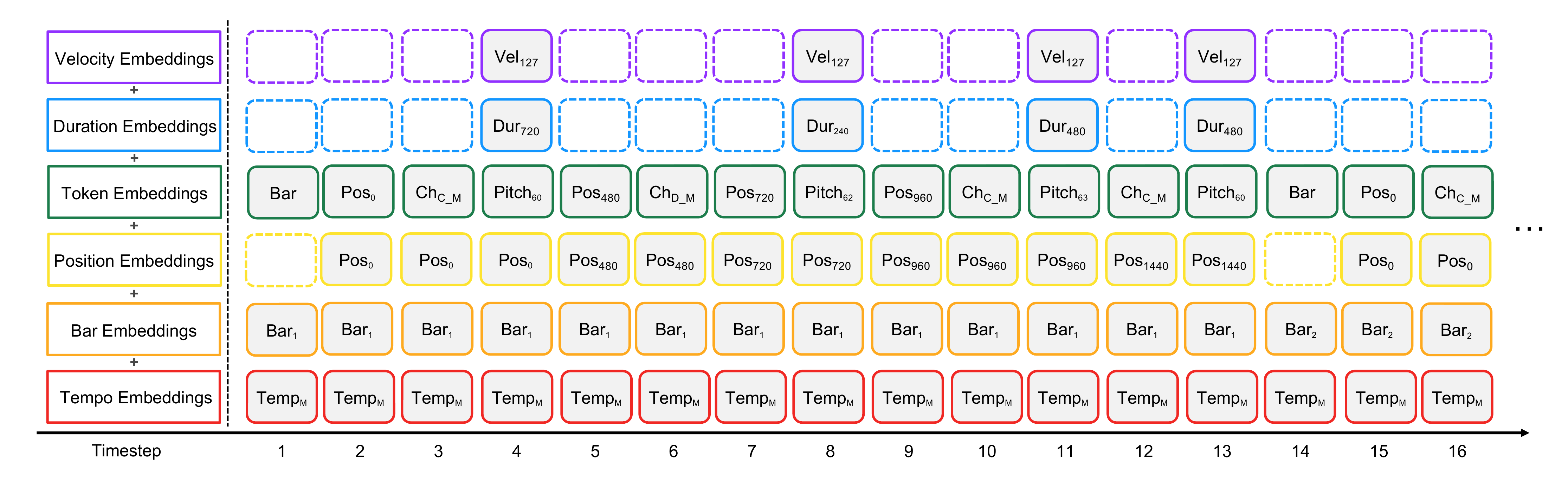}
\caption{\textbf{MeMIDI encoding method for MIDI sequences (example).} The input embedding in each timestep of the MeMIDI event sequence is the sum of the event embeddings, including tempo embedding, bar embedding, position embedding, token embedding, duration embedding, and velocity embedding in this timestep.}

\label{fig_S1}
\end{figure}

\subsection{Additional Supplementary Tables}
\begin{table}[H]
  \begin{center}
\renewcommand\arraystretch{1.25}
  \caption{Subjective evaluation scores of generated melodies based on different melodic skeleton settings in Experiment 1 (mean $\pm$ standard deviation).}
\begin{tabular}{
>{\columncolor[HTML]{FFFFFF}}l 
>{\columncolor[HTML]{FFFFFF}}l 
>{\columncolor[HTML]{FFFFFF}}l 
>{\columncolor[HTML]{FFFFFF}}l 
>{\columncolor[HTML]{FFFFFF}}l 
>{\columncolor[HTML]{FFFFFF}}l 
>{\columncolor[HTML]{FFFFFF}}l }
\hline
{\color[HTML]{000000} \textbf{No.}} & {\color[HTML]{000000} \textbf{Settings}} & {\color[HTML]{000000} \textbf{Rhythm}} & {\color[HTML]{000000} \textbf{Richness}} & {\color[HTML]{000000} \textbf{Structure}} & {\color[HTML]{000000} \textbf{Expectation}} & {\color[HTML]{000000} \textbf{Overall}} \\ \hline
{\color[HTML]{000000} 1}            & {\color[HTML]{000000} Downbeat}          & {\color[HTML]{000000} 2.75 $\pm$  0.96}       & {\color[HTML]{000000} 2.50 $\pm$  0.98}         & {\color[HTML]{000000} 2.63 $\pm$  1.13}          & {\color[HTML]{000000} 2.45 $\pm$  1.03}            & {\color[HTML]{000000} 2.52 $\pm$  1.03}        \\
{\color[HTML]{000000} 2}            & {\color[HTML]{000000} Long Note}         & {\color[HTML]{000000} 2.70 $\pm$ 1.03}       & {\color[HTML]{000000} 2.57 $\pm$  1.17}         & {\color[HTML]{000000} 2.77 $\pm$  1.21}          & {\color[HTML]{000000} 2.47 $\pm$ 1.27}            & {\color[HTML]{000000} 2.67 $\pm$  1.26}        \\
{\color[HTML]{000000} 3}            & {\color[HTML]{000000} Rhythm}            & {\color[HTML]{000000} \textbf{3.02 $\pm$  1.01}}       & {\color[HTML]{000000} \textbf{3.10 $\pm$  0.89}}         & {\color[HTML]{000000} \textbf{2.88 $\pm$ 1.02}}          & {\color[HTML]{000000} \textbf{2.82 $\pm$ 0.81}}            & {\color[HTML]{000000} \textbf{2.93 $\pm$ 0.97}}        \\
{\color[HTML]{000000} 4}            & {\color[HTML]{000000} Tonic}             & {\color[HTML]{000000} 2.95 $\pm$  1.11}       & {\color[HTML]{000000} 2.80 $\pm$  1.09}         & {\color[HTML]{000000} 2.87 $\pm$  0.95}          & {\color[HTML]{000000} 2.65 $\pm$ 1.01}            & {\color[HTML]{000000} 2.80 $\pm$ 1.06}        \\
{\color[HTML]{000000} 5}            & {\color[HTML]{000000} Intersection}      & {\color[HTML]{000000} 2.60 $\pm$  0.85}       & {\color[HTML]{000000} 2.52 $\pm$  0.87}         & {\color[HTML]{000000} 2.53 $\pm$  0.83}          & {\color[HTML]{000000} 2.28 $\pm$  0.98}            & {\color[HTML]{000000} 2.43 $\pm$  0.83}        \\
{\color[HTML]{000000} 6}            & {\color[HTML]{000000} Union}             & {\color[HTML]{000000} 2.95 $\pm$  0.95}       & {\color[HTML]{000000} 2.57 $\pm$  1.05}         & {\color[HTML]{000000} 2.62 $\pm$  1.09}          & {\color[HTML]{000000} 2.43 $\pm$ 1.05}            & {\color[HTML]{000000} 2.60 $\pm$  1.04}        \\
{\color[HTML]{000000} 7}            & {\color[HTML]{000000} Random25\%}        & {\color[HTML]{000000} 2.80 $\pm$  1.11}       & {\color[HTML]{000000} 2.78 $\pm$  0.98}         & {\color[HTML]{000000} 2.67 $\pm$  1.07}          & {\color[HTML]{000000} 2.62 $\pm$  0.95}            & {\color[HTML]{000000} 2.65 $\pm$  1.12}        \\
{\color[HTML]{000000} 8}            & {\color[HTML]{000000} Random50\%}        & {\color[HTML]{000000} 2.92 $\pm$  0.98}       & {\color[HTML]{000000} 2.82 $\pm$  0.89}         & {\color[HTML]{000000} 2.70 $\pm$  1.01}          & {\color[HTML]{000000} 2.68 $\pm$ 1.03}            & {\color[HTML]{000000} 2.73 $\pm$  0.95}        \\
{\color[HTML]{000000} 9}            & {\color[HTML]{000000} Random75\%}        & {\color[HTML]{000000} 2.82 $\pm$  0.95}       & {\color[HTML]{000000} 2.58 $\pm$  1.12}         & {\color[HTML]{000000} 2.50 $\pm$  1.11}          & {\color[HTML]{000000} 2.45 $\pm$ 1.04}            & {\color[HTML]{000000} 2.53 $\pm$ 1.08}        \\ \hline
\end{tabular}
\end{center}
\end{table}

\begin{table}[H]
  \begin{center}
\renewcommand\arraystretch{1.25}
  \caption{One-tailed t-test results between WuYun-RS and other music generation models on the five evaluation metrics in experiment 2.}

\begin{tabular}{
>{\columncolor[HTML]{FFFFFF}}c 
>{\columncolor[HTML]{FFFFFF}}c 
>{\columncolor[HTML]{FFFFFF}}c 
>{\columncolor[HTML]{FFFFFF}}c 
>{\columncolor[HTML]{FFFFFF}}c 
>{\columncolor[HTML]{FFFFFF}}c }
\hline
\textbf{Model} & \multicolumn{1}{c}{\cellcolor[HTML]{FFFFFF}\textbf{Rhythm}} & \multicolumn{1}{c}{\cellcolor[HTML]{FFFFFF}\textbf{Richness}} & \multicolumn{1}{c}{\cellcolor[HTML]{FFFFFF}\textbf{Structure}} & \multicolumn{1}{c}{\cellcolor[HTML]{FFFFFF}\textbf{Expectation}} & \multicolumn{1}{c}{\cellcolor[HTML]{FFFFFF}\textbf{Overall}} \\ \hline
MT             & $3.43\times  10^{-7}$              & $5.37\times 10^{-9}$              & $2.44\times 10^{-12}$              & $6.50\times 10^{-6}$             & $2.21\times 10^{-10}$           \\
PMT            & $2.18\times  10^{-8}$              & $1.21\times 10^{-9}$              & $1.20\times 10^{-6}$              & $2.55\times 10^{-6}$             & $1.13\times 10^{-7}$           \\
MeMIDI         & $1.88\times  10^{-6}$              & $3.09\times 10^{-6}$              & $6.86\times 10^{-7}$              & $2.31\times 10^{-5}$             & $6.29\times 10^{-7}$           \\
CWT            & $3.31\times  10^{-4}$              & $8.47\times 10^{-4}$              & $3.02\times 10^{-4}$              & $2.03\times 10^{-3}$             & $1.69\times 10^{-3}$           \\
Melons         & $4.60\times  10^{-3}$              & $1.54\times 10^{-3}$              & $8.02\times 10^{-4}$              & $2.12\times 10^{-3}$             & $7.41\times 10^{-3}$           \\
WuYun-RRS      & 0.26                             & 0.36                            & 0.42                             & 0.42                           & 0.41                       \\ \hline
\end{tabular}
  {\small
  \begin{tablenotes}
     \item[1]MT, PMT, and CWT stand for Music Transformer, Pop Music Transformer, and Compound Word Transformer, respectively.
   \end{tablenotes}
    }
\end{center}
\end{table}
\end{document}